# Student Engagement in AI-Assisted Complex Problem Solving: A Pilot Study of Human-AI Rubik's Cube Collaboration

Kirk Vancore
Cornell University
kirk.vanacore@gmail.com

Jaclyn Ocumpaugh
University of Houston
jocumpau@cougarnet.uh.edu

Forest Agostinelli
University of South Carolina
kirk.vanacore@gmail.com

Dezhi Wu
University of South Carolina
dezhiwu@cec.sc.edu

Sai Vuruma
University of South Carolina
svaruma@email.sc.edu

Matt Irvin
University of South Carolina
irvinmj@mailbox.sc.edu

## ABSTRACT

Games and puzzles play important pedagogical roles in STEM learning. New AI algorithms that can solve complex problems offer opportunities for scaffolded instruction in puzzle solving. This paper presents the ALLURE system, which uses an AI algorithm (DeepCubeA) to guide students in solving a common first step of the Rubik's Cube (i.e., the white cross). Using data from a pilot study we present preliminary findings about students' behaviors in the system, how these behaviors are associated with STEM skills – including spatial reasoning, critical thinking and algorithmic thinking. We discuss how data from ALLURE can be used in future educational data mining to understand how students benefit from AI assistance and collaboration when solving complex problems.

## 1. INTRODUCTION

Attempts to introduce games into the learning process have varied widely, especially in digital learning technologies. Previous research on learning games has sometimes focused on gamifying learning systems (e.g., adding points to non-game environments), while other research has looked at using immersive game-like environments to teach specific subject material [11,15] . Although games and puzzles were part of the classroom long before the introduction of digital learning platforms, recent advancements in AI allow for potential novel forms of engagement with pedagogically relevant games and puzzles.

Research has shown that games can increase student motivation to persist in the face of difficulty [5,18,21,29], which, in turn, can lead to improved learning [22,3,]. However, there is still much to be learned about how to understand students' behavior within games and how to leverage AI to increase learning and engagement from them. Some researchers have sought to characterize student game play by understanding their goals within the game [3,9,26,38], while more recent studies have used data mining techniques to cluster students into archetypes based on their behaviors using techniques like clustering [23] or epistemic network analysis [39].

In this study, we present a novel system – ALLURE – which uses an AI algorithm (DeepCubeA) to help students engage with and learn from the Rubik's cube. We include a preliminary analysis from a pilot study of ALLURE to understand students' within-system behavioral profiles and how these profiles relate to STEM-related skills, such as spatial reasoning, critical thinking and algorithmic thinking.

## 2. LITERATURE REVIEW
### 2.1 Games and learning

Substantial research has shown that games can improve motivation and learning [5,18,21,29,31]. To date, however, most research within the educational data mining community has been on games that are fully online, as these offer affordances for rich interaction data that are otherwise difficult to obtain from traditional (offline) games.

Within the game-based research, there has also been an effort to characterize different kinds of players. One of the earliest studies to do so was Bartle (1996), who suggested that four player archetypes were common in games known as multi-user dungeons (MUDS [3]). Namely, he characterized players into four groups based on two dimensions, which he likened metaphorically to the suits in a card deck: (1) Diamonds seek to *achieve* treasure; (2) Spades *explore* or dig for information; (3) Hearts are empathetic players who enjoy *socializing;* and (4) Clubs are *killer*s who attack other players.

However, other research suggests that Bartle's categories are insufficient for categorizing archetypes in other types of games [15]. In particular, the fact that Bartle was focused on categorizing players in a highly social, cooperative game platform may make some of those categories less relevant in other kinds of domains, particularly those that are single-player. Moreover, these archetypes were associated with gameplay more generally, and not to learning outcomes in an AI-driven learning environment.

Furthermore, Slater et al, [23] explored the potential for using archetypes to understand student learning behaviors in a single-player game called *Physics Playground* [22]. In their analysis, roughly half of the players could be classified as "explorers," who were relatively good at completing game tasks and appeared to at least be tinkering with the functionality of the game in ways that were likely to improve learning. Another sixth of the players were classified as "achievers," who spent less time interacting with the various parts of the game, due largely to the fact that they were able to successfully complete the tasks with the highest ratings possible in fewer tries. Finally, the remaining third of the players were classified as "disengaged." These students started fewer levels, completed fewer game tasks, and received lower scores for their contributions.

## 2.2 Rubik's cube

This study examines data from students solving the Rubik's cube--a puzzle game designed for teaching spatial reasoning [19]. Originally invented by architect Professor Ernő Rubik, who needed a month to solve the cube for the first time, it became a cultural phenomenon in the 1980s [19]. It then saw a resurgence in popularity after the invention of YouTube, which allowed people to more easily share strategies. Today, the World Cubing Association hosts hundreds of competitions annually and has recorded nearly 100 competitors who have solved the traditional 3x3 cube in less than 5 seconds during an official competition [36].

The Rubik's cube offers several features, making it an ideal puzzle game for understanding how students engage with complex problems. In addition to being a relatively inexpensive object that could be easily distributed at scale, the cube's notorious difficulty level means that students who master it are likely to experience authentic feelings of accomplishment. Such feelings are sometimes considered necessary for improving motivational constructs like resilience or self-efficacy [27].

Moreover, the cube is a good object for teaching computational thinking skills [33]. Solving the cube requires the ability to use several skills related to computational thinking, including the ability to think in multiple levels of abstraction [34], deconstruct the problem into common patterns [35], and formulate potential solutions [33].

Additionally, it offers hands-on learning opportunities that are widely known to be enhanced learning experiences [7], as well as interesting test cases for testing newer, more advanced forms of AI, including those that are using advanced vision technologies. Furthermore, studies have shown that use of the Rubik's cube can help improve key spatial reasoning skills, such as mental rotation [28,29]. Furthermore, [2] found that being taught Rubiks cube solutions was associated with significant increases in problem solving abilities and self-efficacy.

Work to understand the cube and the pattern recognition needed to describe it has not been limited to the enthusiastic community of Cubers who frequent the World Cubing Association Events. Mathematicians have attempted to estimate the "God Number"--or minimum number of moves required to solve the standard 3x3 cube from any given starting state. Rokicki et al. (2014) estimate that number to be just 20 moves. This estimate is down from 26 (Kunkle & Cooperman, 2007), and some scrambles can be solved in fewer moves (see the inverse scramble strategy in Figure 1/Table 1).

Clearly, identifying the kinds of moves that would help you approach the god number for a cube is a skill that is likely to develop over time, and different strategies (such as those shown in Figure 1/Table 1), are reportedly easier to acquire than others [40]. One common technique for solving the cube that is often taught to beginners is called the White Cross method, which advises learners to solve the first layer (side) of the cube in two steps: first by making a white cross and then by using an iterative process to move the white corners into position [20]. Tutorials for this strategy often walk learners through this process, teaching them cube notation as they practice, but most of these tutorials are either static (e.g., [20]) or involve YouTube videos (e.g., [16]), neither of which allow the learner to get advice that is specific to the current state of their cube.

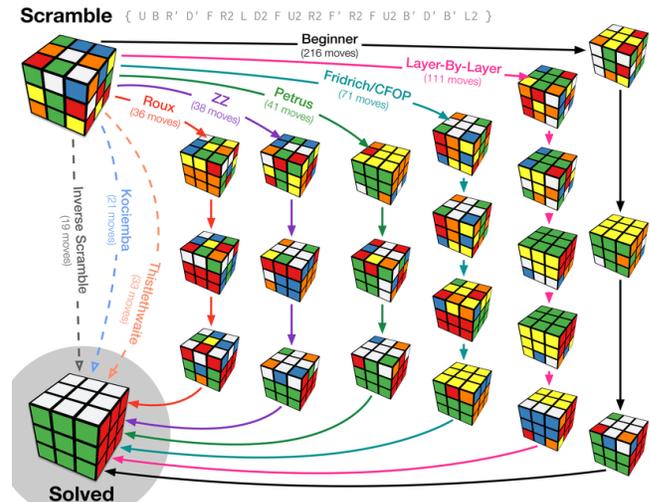

**Figure 1. Multiple strategies for solving a scrambled cube (from [40])**

**Table 1. Number of moves needed to solve the configuration in Figure 1 for any given strategy (from [40])**

| Strategy Name | Moves required |
| --- | --- |
| Beginner: | 216 |
| Layer by Layer: | 111 |
| Fidirich/CFOP: | 71 |
| Petrus: | 41 |
| ZZ: | 28 |
| Roux: | 36 |
| Thistlewaite: | 33 |
| Kociemba: | 21 |
| Inverse Scramble: | 19 |

Developing a system that uses AI to coach students through learning the cube has multiple benefits for students and educational data researchers. For students, tools like these may reduce the barriers to entry-level STEM activities, like the cube, inducing algorithmic thinking while helping them develop spatial reasoning. For educational data mining researchers, it offers opportunities to understand how students engage with AI as they solve complex problems.

## 3. METHODS

This paper reports on a pilot study of an AI-driven system that helps people learn to solve the Rubik's Cube. As described below, we use data from the ALLURE learning system to cluster students into profiles based on their interaction with AI-guided Rubik's cube instructional activities. We then compare these clusters to one another based on how they responded to self-report assessments of cube ability, algorithmic thinking, critical thinking, complex problem-solving skills, and comfort with challenging learning. Finally, we assess how these profiles are associated with differences in their gains in performance on spatial reasoning tasks given before and after ALLURE use.

## 3.1 Context of Study: ALLURE

### 3.1.1 Pilot Study

This study examines pilot data from the ALLURE system [37], which provides AI-guided instructions to students learning to solve a standard (not AI-driven) Rubik's cube. Specifically, we look at data from a pilot test where 31 students at a large university in the SouthEastern United States were asked to complete a white cross task on an AI-driven learning platform called ALLURE, designed to solve a Rubik's Cube. The task completion time ranged from 30 minutes to 2 hours, depending on whether the students were novice or expert Rubik's cube players.

### 3.1.2 Algorithm

The ALLURE system takes a specified goal (e.g., the white cross) and returns an algorithm for achieving that goal, where the algorithm is represented as a directed graph. In this directed graph, nodes represent subgoals, and edges represent algorithms for transforming one subgoal into another. In this context, a subgoal represents a set of Rubik's cube configurations by setting certain stickers to a 7th grey color that indicates that it can be set to any value. For example, a subgoal where all stickers are grey would represent the set of all possible Rubik's cube configurations.

The model for discovering this directed graph is based on two AI algorithms: DeepCubeA [1] and inductive logic programming [6](ILP). DeepCubeA leverages deep reinforcement learning [25] and heuristic search [8] to find a path (a sequence of actions) between any given Rubik's cube state and any given Rubik's cube subgoal. ILP is an AI method that learns logical programs, which can be used to incorporate explainability since logic programs can often be easily converted into a form that humans can understand.

Given a goal the algorithm starts with a graph with that goal as its only node. Then, DeepCubeA is used to find paths to that goal from a variety of other Rubik's cube configurations. The states found are then clustered according to their paths to the goal while also taking symmetry into account. Then, for a cluster closest to the goal, ILP is used to learn a new subgoal, where positive examples are those in the cluster and negative examples are those not in the cluster. This process is then repeated until the subgoals in the graph cover the entire Rubik's cube state space. The final result is then a logic program that can reach the goal from any given state.

### 3.1.3 ALLURE User Interface

Learners interact with the ALLURE system through an AI-guided user interface (Figure 2), which translates the DeepCubeA/ILP logic program into actions a user can take to reach a goal state. This interface is multimodal, providing both animation-based explanations and chat-based instructions designed in line with Explainable AI (XAI) principles to explain AI strategies to the user to decode the black box of AI. The goal of this system is to develop Vygotskian-style scaffolding on increasingly challenging tasks to personalize student learning [37] so that it can support students in both improving spatial reasoning and learning computational thinking skills.

The ALLURE platform has three major function areas (Figure 2). On the left panel, users can engage with a tutorial to help navigate through the platform and explore nine different user case "scenarios" by enabling XAI visualizations via a set of animations, i.e., nine different AI solutions with step-by-step instructions to solve the white cross. In the center, the digital 3D Rubik's cube is positioned for users to move, reset, and use different tools to help solve the white cross. The right panel features a chatbot ALLY, which serves as a virtual AI tutor to assist users in learning different XAI solutions.

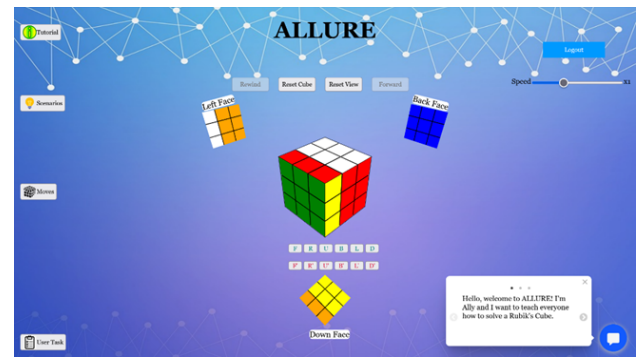

**Figure 2. The ALLURE system interface.**

Given that the possible number of cube states is exceptionally large ($4.3 \times 10^{19}$) [1], so are the number of possible strategies for solving the cube [24]. ALLURE addresses this by providing learners with the opportunity to specify sub-goals, such as reaching the White Cross stage (shown in Figure 3), which is a well-known way to start solutions in Cross, First 2 Layers, Orientation, Permutation (CFOP) method, a typical beginner method.

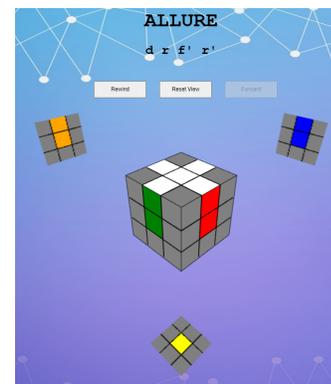

**Figure 3. The White Cross Subgoal**

To test how students might interact with ALLURE, students in this pilot study were encouraged to first complete at least one practice task to familiarize them with the ALLURE platform. (However, they were permitted to skip the practice task if they felt confident enough to do so.) Next, students were presented with two challenge tasks--one involving a simple scramble and one involving a more complex initial configuration. In all tasks, students were asked to produce the White Cross (Figure 3).

Student scaffolding is provided in the practice tasks, where ALLURE provides a digital cube for the students and allows them to manipulate it on the screen to familiarize themselves with the learning system. During practice tasks, students are given the simplest digital Rubik's cube with a configuration in which students can make rotations to understand how the overall ALLURE system works and how they can use different tools built to assist with their problem-solving on the ALLURE platform. As students make rotations to the cube, they also practice interacting with the different functions of the system. These include the commands used to *move* the cube as well as a *reset* function (returning the cube to its initial state for that task), and a *mirror* function (to show hidden faces of the cube).

During these practice tasks include nine different scenarios, which are short tasks of a few cube moves to achieve simple goal states. The tasks are scaffolded by demonstrations form using multi-modal tutorial, that provides both animations and chatbot explanations. These features serve to guide the students through steps determined by DeepCubeA. Forward and rewind functions allow the students to control the presentation of this tutorial.

During challenge tasks, students are asked to complete a more difficult cube state: the White Cross task. During these tasks, students can explore a variety of DeepCubeA generated solutions to get from their cube state to the goal state.

### 3.1.4 Data

The pilot study involves two types of data: log-file data from ALLURE and survey/assessment data taken directly before and after students engaged with the ALLURE system. These data were obtained from the authors [37]. The log file data consists of every action the students took, including each cube movement, each reset the cube to its original state, each time they started, and completed tasks (practice and challenge). For analysis, we aggregated the data at the student level to include the number of practice tasks the students completed (notably, students completed every practice task they started), the number of complex tasks the students started and completed, the number of times student reset the cube to its original position, and the total number of cube moves. We also calculated the percentage of cube moves during each of the tasks (% practice moves, % complex task moves) and the percentage of cube moves students outside of either task (% non-AI guided moves).

Directly before and after the students used ALLURE, they took a survey that included self-report measures on their experience level with the Rubik's Cube. They also took Korkmaz et al.'s [11] computational thinking survey, which includes scales for algorithmic thinking, critical thinking, and confidence in problem-solving. The students also took pre- and post-tests of spatial skills using a classic mental rotation test (MRT; [32]) in which students complete a series of spatial rotation questions that range in various difficulty and complexity levels, where they are presented with a drawing of a 3D object and asked to match it to a drawing where that object has been rotated by selecting the correct answer from a series of distractors.

## 3.2 Analysis

Our analysis involved three steps: first, we identified ALLURE usage profiles using students' aggregated log file data. Second, we evaluated whether the profiles show differences in the surveyed constructs (self-reported Cube Ability, Algorithmic Thinking, Critical Thinking, Complex Problem Solving, Challenging Learning). Third, we evaluate whether these profiles experience different gains in spatial reasoning based on their involvement in this study.

### 3.2.1 Clustering

To understand the patterns of students' behaviors in ALLURE, we used a k-means clustering algorithm – *kmeans* function in R [18]. We clustered based on the following student-level aggregated variables of ALLURE usage, described in **Table 2**. All variables were standardized (z-scored) before being included in the analysis.

**Table 2. ALLURE Interaction Features**

| Feature Name | Operationalization |
|---|---|
| practice tasks | Number of times the students started/completed a practice task |
| challenge tasks started | Number of times the student started a challenge task (solve white cross) |
| challenge tasks completed | Number of times the student completed a challenge task (solve white cross) |
| cube resets | Number of times the student rests the cube during a challenge task |
| cube moves | Number of times the student moved part of the cube during a challenge task |
| % non-AI guided moves | Percentage of moves that a student made outside of any task |
| % practice moves | Percentage of total moves that a student made in practice sessions |
| % challenge tasks moves | Percentage of total moves that a student made on challenge tasks. |

We selected the optimal number of clusters by identifying the "elbow point" in the within-cluster sum of squares (WSS) plot, where the rate of decrease in WSS slows substantially. This is achieved by computing the second-order difference of WSS values and selecting the cluster number where the deceleration is maximized, indicating diminishing returns from increasing the number of clusters.

### 3.2.2 Cluster Differences in Pretest Survey

To compare the differences, we used Bayesian regression models through the *brm* and *emeans* packages in R (Bürkner, 2017). This method allowed us to calculate the probability that each cluster had a different mean on each construct by comparing the posterior distributions of the estimated clusters' means. We used this method instead of a frequentist approach because our small sample would bias *p-value* results towards Type 2 errors (false negatives). Thus, we classified results as being likely different if the probability of the difference being greater than or less than zero is at least .95 and possibly different in that probability was at least .85. Because this is an analysis of a pilot study, which will be replicated, we are interested in investigating possible associations that can be further explored in future work, not only statistically significant ones.

### 3.2.3 Spatial Reasoning Analysis

Finally, we examined the differences in spatial reasoning between the clusters after the use of the ALLURE system. We conducted this analysis by running a Bayesian regression using the *stan_glm* package in r [12]. Once again, we chose a Bayesian method to have more interpreted results of the uncertainty due to the small sample.

We run the regression twice with different outcomes. First, we regress the post-test spatial reasoning on the dummy coded clusters, including the pre-test spatial reasoning as a covariate. Then, we create normalized spatial skill gain delineated in Equation 1, which is used to measure the amount of gain the student made on the skill relative to their original score [41]. This normalized gain score is regressed on the dummy-coded clusters. Together, these models give us an indication of the potential benefits of different types of ALLURE use on spatial reasoning.

$$norm\_gain = \begin{cases} \dfrac{post - pre}{1 - pre} & post > pre \\ \dfrac{post - pre}{pre} & post \leq pre \end{cases} \quad (1)$$

## 4. RESULTS
### 4.1 Descriptive Results

Table 3 presents descriptive statistics of ALLURE usage as well as the survey and assessment variables used through these analyses. On average, students did between one and two practice and one challenge tasks, resetting the cube about eight times. Notably, on average, they start twice as many challenge tasks as they complete. Students also spent much of their time on non-AI guided activities, averaging 87% of cube moves.

**Table 3: Descriptive statistics for both usage and survey/assessment variables**

|  | Mean | Min | Max | SD |
|---|---|---|---|---|
| **ALLURE Usage Variables** | | | | |
| practice tasks | 1.45 | 0.00 | 9.00 | 2.31 |
| challenge tasks started | 1.94 | 0.00 | 8.00 | 2.48 |
| challenge tasks completed | 1.06 | 0.00 | 3.00 | 1.18 |
| cube resets | 8.23 | 0.00 | 38.00 | 7.94 |
| cube moves | 119.65 | 5.00 | 368.00 | 72.50 |
| % non-AI guided moves | 0.87 | 0.24 | 1.00 | 0.20 |
| % practice moves | 0.02 | 0.00 | 0.18 | 0.04 |
| % challenge moves | 0.11 | 0.00 | 0.76 | 0.19 |
| **Survey/Assessment Variables** | | | | |
| Self Report Cube Ability | 1.81 | 1.00 | 4.00 | 1.01 |
| Algorithmic Thinking | 14.62 | 5.00 | 23.00 | 4.80 |
| Critical Thinking | 18.84 | 14.00 | 24.00 | 2.44 |
| Complex Problem Solving | 3.48 | 2.00 | 5.00 | 0.77 |
| Challenging Learning | 4.42 | 3.00 | 5.00 | 0.56 |
| Pretest Spatial Reasoning | 0.58 | 0.00 | 1.00 | 0.30 |
| Posttest Spatial Reasoning | 0.63 | 0.08 | 1.00 | 0.25 |
| Normalized Spatial Reasoning Gain | 0.16 | -0.89 | 1.00 | 0.44 |

### 4.2 Clustering Results

The elbow method suggested a three-cluster model, the results of which are in Table 4. We labeled these profiles *Challengers*, *Explorers*, and *Emerging Strategists.* These labels reflect categories that are common in the literature, where students often pattern differently depending on goals and prior knowledge (Bartle, 1996; Zambrano et al., in press).

**Table 4. Clustering results showing z-scores of the ALLURE Activities**

| Cluster | Challengers | Explorers | Emerging Strategists |
|---|---|---|---|
| practice tasks | 0.02 | 1.16 | -0.55 |
| challenge tasks started | 1.10 | 0.63 | -0.69 |
| challenge tasks completed | 0.93 | 1.00 | -0.80 |
| cube resets | 0.12 | -0.60 | 0.24 |
| cube moves | 0.60 | -0.16 | -0.14 |
| % non-AI guided moves | -1.74 | -0.01 | 0.62 |
| % practice moves | -0.16 | 1.17 | -0.49 |
| % challenge moves | 1.84 | -0.22 | -0.54 |

The *Challengers* tend to avoid the practice elements of ALLURE and spend the majority of their moves in the challenge tasks. They also make the least amount of moves outside of AI-guided tasks. They reset their cubes slightly more than average and make more cube moves overall than the other profiles. This profile suggests high engagement with ALLURE, focused on accomplishing the challenging elements of the systems, without relying too heavily on the scaffolding they would have received during the practice tasks.

The *Explorers* behavior is different. They spend more time completing practice tasks than the other clusters, as evidenced by both the higher number of practice tasks started and the percentage of moves that they've completed that occur within those tasks. Interestingly, they start fewer challenge tasks than *Challengers*, but also complete a relatively high number (essentially equivalent to the numbers seen for the *Challengers*). They also reset the cube less frequently than average, possibly because they have benefited from practice and need less trial and error. They spend most of their time (as measured in cube moves) in practice and an average amount of time using the ALLURE with no AI guidance.

The *Emerging Strategists* show behaviors that appear more random than the other profiles. They make as many moves as the *Explorers* but they seem to be avoiding help, as they make the least use of ALLURE's AI guidance, evidenced by low engagement in either the practice or the challenge tasks. Thus they are least likely to complete the challenge tasks. They reset the cube more often than average, suggesting they were the least unlikely to complete any challenge tasks.

### 4.3 Profile Differences in Pretest Survey

Table 5 presents how the three student behavioral profiles differed in terms of the five surveyed constructs. Results show low certainty of differences between profiles on self-reported abilities of algorithmic thinking or of critical thinking skills, nor of their self-reported ability with the Rubik's cube. However, there were likely differences between profiles on the complex problem solving measure and on the challenging learning measure.

**Table 5. Profile Differences on Survey Items with uncertainty estimates (P(0), P(>0))**

| Variable | Explorers vs. Challengers | Explorers vs. Emerging Strategists | Challengers vs. Emerging Strategists |
|---|---|---|---|
| Self Report Cube Ability | 0.06 (0.461, 0.539) | 0.11 (0.402, 0.598) | 0.06 (0.45, 0.55) |
| Algorithmic Thinking | 0.29 (0.302, 0.699) | -0.09 (0.584, 0.416) | -0.38 (0.785, 0.215) |
| Critical Thinking | -0.04 (0.528, 0.472) | -0.46 (0.842, 0.158) | -0.43 (0.8, 0.2) |
| Complex Problem Solving | **-0.5 (0.882, 0.117)** | **-0.7 (0.98, 0.02)** | -0.2 (0.712, 0.288) |
| Challenging Learning | **0.79 (0.004, 0.997)** | 0.10 (0.323, 0.677) | **-0.70 (0.996, 0.004)** |

Bolded differences are either likely (P(<|>0) > .95 or suggestive P(<|>0) > .85 )

On average, the *Challengers* rated themselves higher in complex problem-solving (e.g., "I feel confident when working on complex problems") than the *Explorers*. This makes sense as these students spent little time (in terms of moves) on the practice tasks, perhaps because they were confident that they could learn the challenging tasks without the scaffolded instruction of the practice tasks. *Emerging Strategists* also rated themselves more highly on complex problem solving than the *Explorers*, despite not engaging with the challenging tasks. This may suggest overconfidence.

However, both *Explorers* and *Emerging Strategists* rated themselves higher for challenging learning (e.g., "I am willing to learn challenging things.") than the *Challengers*. This is particularly interesting as the *Challengers* seemed to spend more time (measured in cube moves) on the challenge tasks than other *Explorers*. However *Explorers* completed a similar number of challenging tasks. Thus, it could be that the *Explorers* are willing to embrace challenges, but recognize practice as a part of the process. Once again, *Emerging Strategists'* higher self rating on challenging learning than those who completed the challenge task – *Challengers* – suggests a lack of metacognitive awareness for this group..

## 4.4 Spatial Reasoning Gains Analysis

Finally, because ALLURE was designed to help students develop spatial reasoning, we analyzed the difference between clusters on the post-ALLURE system using spatial reasoning and their normalized spatial skill gains using two regression models. The parameter estimates of these models are presented in Table L. The *Explorers* were used as the reference category in both regressions. These results show that those in the *Challenger* cluster likely had higher post-test spatial reasoning scores (Estimate = 1.63, CI[0.31, 2.94]) and normalized spatial reasoning Gains scores (Estimate = 0.51, CI[-0.04.0.9]).

**Table 6. Parameter Estimates and 95% credible intervals (CI) from clustered regressed on spatial reasoning test measures**

| | Post Test Scores | | Normalized Gains | |
|---|---|---|---|---|
| Variable | Estimates | CI | Estimates | CI |
| Intercept | 1.48 | 0.26 – 2.65 | -0.01 | -0.31 – 0.29 |
| Challengers | **1.63** | **0.31 – 2.94** | **0.51** | **0.04 – 0.96** |
| Emerging Strategists | 0.67 | -0.38 – 1.74 | 0.11 | -0.25 – 0.49 |
| Pre spatial reasoning | **0.46** | **0.20 – 0.72** | | |

Estimates are bolded when CI does not span zero (the equivalent of statistical significance in frequentist statistics)

## 5. DISCUSSION

This study presents preliminary findings to understand the different ways in which students collaborate with AI-guided systems to solve complex puzzles. We also find some promising results connecting patterns of behavioral engagement (i.e., student profiles) with the AI-guided user interface (ALLURE) and gains of spatial reasoning. Overall, this pilot study suggests that systems like ALLURE may provide two benefits to the educational data mining community: providing opportunities to study how students can collaborate with AI in complex problem solving while helping students develop important STEM skills.

Notably, we find three behavioral profiles in this study which overlap with some previous classifications of gameplay [3,15], but they best aline with Salter et al.; "achievers", "explorers", and "disengaged" [23] align while with our *Challengers, Explores, Emerging Strategists* (although we have chosen a more asset based description).

Furthermore, they have distinct interaction patterns with ALLURE as well as notable differences between self-perceptions of complex problem solving abilities and willingness to learn from challenging things. For example, *Challengers* make the greatest learning gains on special reasoning and spend most of their time engaging with the AI-guided challenging tasks, making many moves and resetting the cube often. They see themselves as being good at complex problem solving. This suggests that time with challenging AI-guided puzzle solving helped them grow in terms of spatial reasoning. In contrast, the interactions that *Explorers* and *Emerging Strategists* have with the ALLURE system are not associated with gains in spatial reasoning. Since these students have different behavioral patterns, this opens questions about how to design systems like ALLURE to meet the learning needs of students with a wide range of engagement tendencies.

The relationships between students' usage patterns and perceived STEM-related skills – how students characterized their critical thinking, and algorithmic thinking – suggest that students may respond differently to systems like ALLURE based on how they view their STEM abilities. Although complex problem solving and challenging learning scenarios are not peculiar to STEM fields, they are important for STEM learning and occupational success. It is particularly interesting that *Emerging Strategists*, who were least likely to successfully complete the challenge tasks, reported higher confidence in problem-solving, than *Challengers* who were more successful in ALLURE. Findings like these suggest that students may engage and benefit less from AI collaboration if they are overconfident in their skills.

## 6. Future Directions

Given these preliminary findings, future iterations of ALLURE could incorporate adaptive AI models that adjust guidance in response to learner behaviors, similar to those used in intelligent tutoring systems. For instance, AI-driven real-time feedback on inefficient move sequences could better support students who underutilize guidance. Additionally, enhancing explainability in AI feedback—such as justifying recommended moves—may help students develop algorithmic thinking, an approach supported by computational thinking research.

Although the current study aggregated data at the student level, ALLURE offers the possibility of mining rich interaction data in a complex puzzle alongside AI-giaince and feedback. Systems like ALLURE present possibilities for exploring the details of exactly how students engage with AI instructions. Future analyses could examine patterns of cube moves based on structures and explore how students follow and diverge from AI-guidance.

## 7. ACKNOWLEDGMENTS